\documentclass[preprint,12pt]{elsarticle}
\biboptions{numbers,sort&compress} 

\usepackage{fullpage}
\usepackage{graphicx}
\usepackage[center]{caption}
\usepackage{subcaption}
\usepackage{multirow}
\usepackage{hhline}
\usepackage{gensymb}
\usepackage{amsmath}
\usepackage{amssymb}
\usepackage{rotating}
\usepackage{float}
\usepackage{xcolor}
\usepackage{adjustbox}
\usepackage{import} 
\usepackage{array}
\usepackage{import}
\usepackage{xifthen}
\usepackage{pdfpages}
\usepackage{transparent}
\usepackage[hidelinks]{hyperref}
\usepackage{pstool}

\usepackage{paralist}
\restylefloat{table}

\newcommand{\etac}[0] {\ensuremath{\eta_{critical}}\ }

\newcommand{\etatau}[0] {\ensuremath{\eta_{\tau}}\ }
\newcommand{\taupg}[0] {\ensuremath{\tau_{pg}}\ }

\newcommand{\qp}[0] {\ensuremath{q_p}\ }
\newcommand{\qpl}[0] {\ensuremath{q_{pl}\ }}
\newcommand{\qps}[0] {\ensuremath{q_{ps}}\ }

\usepackage{framed}
\usepackage[justification=centering,singlelinecheck=false]{caption}
\journal{Applied Energy}

\begin{document}

\begin{frontmatter}



\title{A Physics-based Scaling of the Charging Rate in Latent Heat Thermal Energy Storage Devices}


\author[1]{Kedar Prashant Shete \corref{cor1}}
\ead{kshete@umass.edu, kedar.kshete@gmail.com}

\author[2]{S.~M.~de Bruyn Kops}
\ead{debk@umass.edu}

\author[3]{Dragoljub (Beka) Kosanovic}
\ead{kosanovic@umass.edu}

\cortext[cor1]{Corresponding author}

\address[1]{Mechanical and Industrial Engineering,
	219 Engineering Laboratory,
	University of Massachusetts,
	160 Governors Drive,
	Amherst, MA 01003-2210}

\begin{abstract}
Thermal energy storage (TES) is increasingly recognized as an essential
component of efficient Combined Heat and Power (CHP), Concentrated Solar Power
(CSP), Heating Ventilation and Air Conditioning (HVAC), and refrigeration as
it reduces peak demand while helping to manage intermittent availability of
energy (e.g., from solar or wind).  Latent Heat Thermal Energy Storage (LHTES)
is a viable option because of its high energy storage density. Parametric
analysis of LHTES heat exchangers have been focused on obtaining data with
laminar flow in the phase changing fluid and then fitting a functional form,
such as a power law or polynomial, to those data. While this approach can
produce an accurate correlation applicable within the range of data used for
its creation, it does not reveal details about the underlying physics. In this
paper we present a parametric framework to analyze LHTES devices by
identifying all relevant fluid parameters and corresponding dimensionless
numbers. We present 64 simulations of an LHTES device using the finite volume
method at four values of the Grashof, Prandtl and Reynolds numbers in the
phase change material (PCM) and heat transfer fluid (HTF). We observe that
with sufficient energy available in the HTF, the effects of the HTF Reynolds
number and Prandtl number on the heat transfer rate are negligible. Under
these conditions, we propose a time scale for the variation of energy stored
(or melt fraction) of the LHTES device based on the Fourier number($Fo$),
Grashof number($Gr_p$) and Prandtl number($Pr_p$) and observe a $Gr_p^1$ and
$Pr_p^{(1/3)}$ dependency. We also identify two distinct regions in the
variation of the melt fraction with time, namely, the linear and the
asymptotic region. The linear region is characterized by constant and
high heat transfer rates, making it the relevant region for operating an
energy storage device. We also predict the critical value of the
melt fraction at the transition between the two regions. From these analyses, 
we draw some conclusions regarding the design procedure for
LHTES devices.
\end{abstract}


%
%
%
\end{frontmatter}
%
%
%
\section{Introduction}
\label{sec:Intro}
Thermal energy storage (TES) is increasingly recognized as an essential
component of efficient combined heat and power (CHP), concentrated solar power
(CSP), heating ventilation and air conditioning (HVAC), and refrigeration as
it reduces peak demand while helping to manage intermittent availability of
energy (e.g., from solar or wind).  As discussed in more detail below, it has
the potential to reduce energy consumption and reduce pollution generation by
making existing technologies more efficient and by enabling
the integration of renewable energy sources with minimum energy curtailment.

Given the thermo-physical properties of a heat storage material, it is
straightforward to compute the amount of that material required to store a
given amount of heat.  The challenge is in designing a physical device that
enables sufficiently high heat transfer rate for a practical system.  If, for
example, a TES is to be coupled with a CHP plant, the TES must be able to
store and release heat at the time scale of the transients in the CHP system.
Designing a TES system to meet this requirement is difficult because a very
large number of parameters affect the heat transfer rate including the
properties of the working fluids, the fluid dynamical regimes of those fluids
when the system is operating, the geometry of the heat exchanger and
storage device, and the operating conditions for the entire system.  In this
paper, we present an approach for dealing with this complexity that consists of
systematically defining the relevant dimensionless parameters and then writing
the relationships between these parameters based on physical understanding
derived from theory and from the literature about TES systems.
Of course this approach is not unique to this paper, but we apply it the
specific case of latent heat thermal energy storage (LHTES) to demonstrate how
the approach can introduce physical understanding into relationships between
parameters that have typically been studied empirically and, thereby, simplify
the overall design process.

In the remainder of this section we review the motivation for studying TES
and, in particular, applications that significantly benefit from LHTES. We
then review some of the fundamental studies in LHTES that provide the physical
understanding necessary for our approach.  In
\S\ref{sec:parametrizing_the_problem} the LHTES problem is defined in terms of
dimensional parameters and dimensionless groups of parameters.  Our numerical
methods and simulation validations are presented in \S\ref{sec:numerical} and
\S\ref{sec:Simulations}.  We demonstrate our methodology in
\S\ref{sec:regions} and \S\ref{sec:Results} to show that what might appear to be empirical
relationships between variables are consistent, when appropriately
parameterized, with basic theories of heat and mass transfer.  Some conclusions regarding the design of LHTES devices are
drawn in \S\ref{sec:conclusions}

\subsection{Applications for Thermal Energy Storage}
Applications that are being improved significantly with thermal energy storage
include concentrating solar power (CSP) plants; \citet{denholm2010} report
round-trip efficiencies close to 100\% when energy from CSP's is stored as thermal energy rather than electrical energy.  They also report that ``cold
storage'' enables extremely high efficiency of cooling systems by shifting
demand to off-peak hours.  \citet{nithyanandam13}, based on their study of
charging and discharging cycles of a LHTES heat exchanger, emphasize the
importance of LHTES for the effective functioning of CSP. Performance of
cogeneration power plants also improves when they are combined with thermal
energy storage \citep{hu2017,mcdaniel16}.  \citet{venkitaraj18} investigate experimentally the use of
nano-particle enhanced LHTES for waste heat recovery from IC engines and
observe up to 18\% increase in the energy savings.

In addition to improvements in energy efficiency, thermal energy storage can
reduce emission of pollutants.  For example, \citet{li17} calculate the
effect of a LHTES system used to recover waste heat from a heavy duty diesel
engine and conclude a potential 40\% improvement in engine warm up time
during which the engine produces suboptimal emissions.  \citet{arbabzadeh2019}
report the huge potential impact of energy storage on decarbonization of
electricity production by allowing electricity usage for heating and cooling
to be synchronized with when renewable energy is available.  Specifically,
they conclude that, for the state of California, thermal energy storage can
result in an 18\% reduction in carbon dioxide emission and a 21\% reduction in
renewable energy curtailment, that is, the reduction of output of a renewable
resource below what it could have otherwise produced.

\subsection{Characteristics of LHTES}
Thermal energy storage can be 
classified
into three major types: sensible heat storage, latent heat storage and
thermochemical energy storage. For the applications discussed in the preceding
paragraphs, a desired 
characteristics of TES include: 
\begin{inparaitem}
	\item High volumetric energy storage density
	\item Heat recovery at constant temperature
	\item Low cost
	\item Fast heat transfer rate.
\end{inparaitem}
LHTES has inherent advantages over other TES systems with respect to high
storage density and heat transfer at constant temperature.  High storage
density, in turn, tends to lead to lower cost.  Thus, LHTES would appear to be
a very attractive option for improving the energy efficiency and reducing
emmisions of a variety of types of power plants and engines.  Indeed,
\citet{mongibello2014} study two different types of thermal energy storage
for residential micro-CHP systems and conclude that LHTES is preferred
over sensible energy storage (such as hot water) in terms of cost and
size.   They
also conclude that further analysis should be made, including of the long-term
performance and degradation of these systems over time, in order to assess the
convenience of using them for thermal energy storage. \citet{johar2017}
implement a LHTES system within a micro-CHP plant and shows LHTES can be a
viable option.  They note, though, that improved design procedures and
performance modeling of phase change heat exchangers are essential.

The last characteristic in the list above, fast heat transfer rate, is the
motivation for the research reported in this paper.  Heat transfer rate
is determined primarily by the fluid dynamics and geometry of the heat exchanger
rather than specifically by the storage mechanism, with turbulent flow over
large surface areas leading to high heat transfer rates.  As reviewed in
\S\ref{sec:heatTransferRate}, understanding the heat transfer rates in the context of
flow of phase changing materials is important for developing practical LHTES systems.

\subsection{Heat transfer rate}
\label{sec:heatTransferRate}
Given the latent heat of fusion of a phase changing material (PCM), it is
relatively simple to calculate the amount of PCM that a LHTES system needs in
order to store a specific amount of energy.  The challenge is in designing a
system with the required {\em heat transfer rate}, which as evident from previous studies depends on a number of geometrical, material and operating parameters.  Given the
complexity of the problem, it is common for individual research studies to
focus on a subset of the parameters affecting heat transfer rate.  An
important first step is to begin with simplified governing equations for heat
transfer, for example, neglecting convective heat transfer in the PCM
\citep{cao1991,cao1992,yimer97,bechiri2015,teamah2016}. Natural convection,
however, is a key component of accurately modeling energy storage rates
\citep{bechiri2015}.

\subsubsection{Geometry}
Heat exchange geometry is a crucial factor affecting the heat transfer
rate of LHTES.  Geometry parameters that have been studied include 
the inner and outer diameters in an annular geometry with PCM in the 
annulus and HTF in the inner pipe \citep{cao1991,kalapala2018}, HTF pipe 
wall thickness \citep{cao1992} and diameter of the HTF pipe \citep{yimer97}.
Adding fins in the PCM has been shown to improve charging rates, stored 
energy and melting front depth \citep{yimer97,kalapala2018,bhagat2018}.
\citet{bhagat2018} conduct an optimization study of fin height, fin thickness and 
number of fins using ANSYS Fluent and laboratory scale heat exchanger data and conclude that 
for a given percentage of fin material/metal inside the heat exchanger, a higher 
number of thinner fins lead to better heat transfer. The overall configuration of 
the LHTES is also an important factor, and various configurations including a single HTF pipe 
inside an annular PCM container, multiple HTF tubes inside a PCM pipe, PCM modules 
floating inside an HTF container and direct contact between HTF and PCM have been studied 
\citep{gasia2017}. The orientation of the device also affects its performance and has 
been studied by \citet{kalapala2018}.

\subsubsection{Thermophysical properties}
The thermophysical properties of the HTF and PCM such as thermal conductivity
and specific heat capacity are also important parameters affecting the
performance of LHTES \citep{cao1991,cao1992,yimer97,farid2004,gasia2017}.
\citet{gasia2017} conclude that an increase in specific heat capacity of HTF
of 4.9 times and in thermal conductivity of HTF of 3 times improves the
charging times by 44 \%.  \citet{farid2004} note the importance of material
properties by observing that materials such as paraffins have moderate energy
storage density and low cost, but also have low thermal conductivity, which
affects their utility as energy storage materials.  Hydrated salts, on the
other hand have larger thermal conductivity and large energy storage capacity,
but their use is affected by other material properties like supercooling and
phase segregation. They conclude that the melting point is the most important
characteristic in selecting a phase change material and point out the
importance of creating materials that have an adjustable melting point.

\subsubsection{System operating parameters}
System operating parameters such HTF mass flow rate and temperature have a
dominant effect on the LHTES performance because it is the HTF that determines
the maximum rate at which energy can be exchanged with the PCM.  The effect of
HTF mass flow rate and HTF temperature has been studied by a number of authors
\citep{kalapala2018,teamah2016,bechiri2015}. For example, a study
conducted in terms of dimensionless parameters is that of \citet{teamah2016},
which was a parametric numerical finite difference analysis of total heat
transfer gain in an cylindrical tank with encapsulated PCM. The parameters
they studied are the HTF Reynolds number in the range $20<Re<4000$, Stefan
number in the range $0.1<Ste<0.4$, $0.2<(\rho C_{p})^*<0.8$ where $(\rho
C_{p})^*$ is the ratio of effective thermal capacity (Density$\times$Specific
Heat Capacity) of the PCM to HTF, $0.2<\theta_{m}<0.8$ where $\theta_{m}$ is
the ratio of the difference between the PCM melting and HTF inlet temperature
to the difference between the HTF inlet temperature and the starting
temperature of the system, and the Fourier Number ${Fo}$ which is
non-dimensional charging time.   They obtained a dependency of ${Fo}
Re^{0.8} \theta_{m}$ and $ Ste(\rho C_{p})^*$ for the total energy gain and
concluded that the dependency of $Re^{0.8}$ originates from the turbulent
convection coefficient correlation used within their finite difference
calculation. Understanding the effect of individual parameters on the
performance and quantifying their importance relative to other parameters will
greatly support the design process for LHTES devices \citep{farid2004}. It is
advantageous to have dimensionless results instead of purely experimental data
pertaining to just one device \citep{bechiri2015}.

\subsubsection{Need for further research and our contribution}
From the foregoing review, it is apparent that the foundation has been laid
for understanding the individual factors affecting heat transfer rate in LHTES
systems.
Less progress has been made on combining these individual factors to
form a complete set of relevant dimensionless parameters suitable for robust
modeling and design guidance for creating LHTES systems having sufficiently high heat
transfer rates for commercial applications.  We begin our study in
\S\ref{sec:parametrizing_the_problem} by identifying the physical parameters
and the corresponding dimensionless numbers and discuss the physical
importance of each for fast heat transfer. In \S\ref{sec:numerical} and
\S\ref{sec:Simulations}, we discuss the equations used and validation of our
simulations. In \S\ref{sec:regions} and \S\ref{sec:Results} we demonstrate the utility of this
parameter set for understanding and modeling, based on numerical simulations,
the physical mechanisms controlling the heat transfer rate. For specificity,
we focus on the effects of four important parameters: HTF inlet velocity, HTF
inlet temperature, HTF thermal conductivity and PCM thermal conductivity on
the heat transfer rate and thermal charging time. In \S\ref{sec:lin_behave} and
\S\ref{sec:exp_behave}, we identify two distinct regions in the heat transfer
rate that explain the reduction in heat transfer and identify a critical
percentage of melting that separates these regions. In \S\ref{sec:Results}, we examine the
underlying convection physics and propose scaling laws for heat transfer rate
as a function of the Reynolds number in HTF, Grashof number in PCM and Prandtl
numbers in both the HTF and PCM. Some conclusions about the
scaling obtained and the reason causing these regions are presented in
\S\ref{sec:conclusions}.

\begin{table}
	\label{tab:nomenclature}
	\begin{framed}
		\begin{tabular}{c>{\raggedright\arraybackslash}p{0.35\textwidth}c>{\raggedright\arraybackslash}p{0.35\textwidth}}
	\multicolumn{4}{l}{\bf Nomenclature} \\
	Symbol & Description 	& Subscript & Description \\
	$\eta$ & Melt fraction 	&	$f$		&	Heat transfer fluid (HTF)		\\
	$\rho$	&	Density		&	$p$		&	Phase change material (PCM)		\\
	$\mu$	&	Viscosity	&	$t$		&	HTF tube		\\
	$\beta$	&	Volumetric expansion coefficient & 	$c$	&	PCM container	\\
	$T$		&	Temperature & 	$i$	&	Inner	\\
	$u$		&	Velocity 	& 	$o$	&	Outer	\\
	$M$		&	Mass 		& 	$in$	&	Inlet	\\
	$\alpha$ &	Thermal diffusivity & 	$mean$	&	Mean	\\
	$\nu$ 	&	Kinematic viscosity & 	$fr$	&	HTF Reynolds number	\\
	$q$ 	&	Heat transfer rate into control volume & 	$fp$	&	HTF Prandtl number	\\
	$h$ 	&	Mean heat transfer coefficient & 	$pg$	&	PCM Grashof number	\\
	$D$ 	&	Diameter 	& 	$pp$	&	PCM Prandtl number	\\
	$A_{mush}$ 	&	Mushy zone constant 	& 	$\tau$	& Dimensionless time	\\
	$\lambda$ 	&	Liquid fraction 	& 	-	&	-	\\
	$\tau$ 	&	Generic dimensionless time 	& 	-	&	-	\\
\end{tabular}%
	\end{framed}	
\end{table}

\section{Parametrizing the problem}
\label{sec:parametrizing_the_problem}
A variety of configurations exist for LHTES systems, but they have certain common elements. Typical LHTES devices consist of a heat
exchanger with a heat transfer fluid (HTF), such as oil, pumped across one
side of a solid interface and a PCM driven by
natural convection on the other side.  Starting from the solid state in the
PCM, introduction of heat to the system via the HTF melts some of the PCM and
buoyancy begins to drive flow. 
Three factors quantify the practical performace of an energy storage/LHTES device:
the charging rate, the discharging rate and the storage capacity.  In a LHTES
device, the storage capacity is very simple to predict because it is directly
proportional to the mass of the of PCM in the system.  The charging and
discharging rates are more difficult to predict because, as reviewed in the
previous section, they depend on the geometry of the heat transfer surface, the
thermophysical propoerties of the fluids, and the operating conditions of the
entire system.  Here we consider only the charging rate because, while the
discharging rate may be different, the approach to parameterizing the modeling
both rates is the same.

A common approach to modeling the charging rate is to fit an assumed function
to experimental or numerical data.  For example, \citet{jyotirmay15} use
polynomial regression to describe the melting time as a function of the
Reynolds number in the HTF, the Stefan number of the PCM and the ratio of
initial temperature of the PCM and inlet temperature of the HTF. \citet{diarce18} assume a
product of power-law relationship to fit the Fourier number as a function of the Biot
number, the Stefan number and two dimensionless temperature constants.  This
approach can produce effective correlations over the range of data used to
produce them but offer limited physical insight to enable predicting
the heat transfer rate outside the range that was measured.

\subsection{Physics-infused correlations}
In commercial applications, the geometry of the heat exchanger is such that
turbulent flow of the PCM can be expected unless the melted fraction is
extremely small. Turbulent flow studies in LHTES systems are limited by the
practical size of laboratory experiments and current limitations in computing
capability.  Therefore, we procede using a physics-based approach to
hypothesize the correct functional forms for the relationships between
dimensionless flow parameters.  This approach begins with identifying the
dimensional system parameters expected to be important for the performance of
LHTES systems.
These are tabulated in Table \ref{tab:paralist}.
\begin{table}
	\centering
	\caption{Dimensional parameters that affect heat transfer rate of LHTES devices. Boxed parameters have been used for the Buckingham Pi analysis.}
	\begin{tabular}{>{\raggedright\arraybackslash}p{0.3\textwidth}cccc}
		Properties & HTF   & PCM   & HTF Tube & PCM Container \\
		\hline\\
		Density & {$\rho_f$} & {$\rho_p$} & $\rho_t$ & $\rho_c$ \\
		Specific Heat Capacity & {$Cp_f$} & {$Cp_p$} & $Cp_t$ & $Cp_c$ \\
		Viscosity & {$\mu_f$} & {$\mu_p$} & -     & - \\
		Thermal Conductivity & \fbox{$k_f$} & \fbox{$k_p$} & \fbox{$k_t$} & $k_c$ \\
		Volumetric Expansion Coefficient  & {$\beta_f$} & \fbox{$\beta_p$} & $\beta_t$ & $\beta_c$ \\
		Inlet Temperature & {$T_{in}$} & -     & -     & - \\
		Initial Temperature & \multicolumn{4}{c}{ $T_i$ } \\
		Freezing Temperature & -     & {$T_{solidus}$} & -     & - \\
		Melting Temperature & -     & {$T_{liquidus}$} & -     & - \\
		Latent Heat Capacity	&	-	&	\fbox{$L$}	&	-	&	-	\\
		Time  			& 	\multicolumn{4}{c}{ \fbox{$t$} }   			\\  
		Time for Solidification with Under-cooling & -	& \fbox{$\Delta t_s$} & - & - \\
		Average Inlet Velocity/Average Velocity & \fbox{$u_f$}	& - & - & - \\
		Length & -	& - & \fbox{$l_t$} & \fbox{$l_c$} \\
		Initial Mass & $M_f$ & $M_p$ & $M_t$ & $M_c$ \\
		Diameters	&	- &	- & \fbox{$D_t$}	& \fbox{$D_{ci}, D_{co}$} \\
		Container to Fluid Interface Area	&	$A_f$ &	$A_p$ &  -	& - \\
		\\
		\multicolumn{4}{l}{Derived parameters}\\
		\hline\\
		Mean Surface Temperature ($\int_{A}TdA/A$) & - & - & - & \fbox{$T_{mci}$} \\
		Mean Heat Transfer Coefficient ($h$) & \fbox{$h_f$} & \fbox{$h_p$} & - & - \\
		Mean Melting Temperature ($(T_{solidus} + T_{liquidus})/2$) & - & \fbox{$T_{mean}$} & - & - \\
	\end{tabular}%
	\label{tab:paralist}%
\end{table}%

\begin{table}
	\centering
	\caption{Other derived parameters}
	\begin{tabular}{>{\raggedright\arraybackslash}p{0.3\textwidth}cccc}
		Properties & HTF   & PCM   & HTF Tube & PCM Container \\
		\hline\\
		Thermal Diffusivity ($k/\rho C_p$) & {$\alpha_f$} & {$\alpha_p$} & - & - \\
		Kinematic Viscosity (${\mu}/{\rho}$) & {$\nu_f$} & {$\nu_p$} & - & - \\
		Mean volume temperature ($\int_{V}TdV/V$) & {$T_{f}$} & {$T_{p}$} & - & - \\
		Heat transfer rate out of/into control volume & $q_f$	& $q_p$	& $q_t$	& $q_c$ \\
	\end{tabular}%
	\label{tab:other_paralist}%
\end{table}%

We note that the number of parameters affecting LHTES system performance is
extremely large.  For example, the macroscopic geometry of the device can be quite
complicated and, e.g., microsopic geometry of the heat transfer
surfaces is a topic unto itself.  Here we have assumed an annular geometry of
smooth-walled tubes with the HTF in the inner tube and PCM in the annulus.

\subsubsection{Application of Buckingham Pi theorem}
Given the very large number of parameters in Table \ref{tab:paralist} and are
narrowing of the focus of this paper to a simple geometry, we procede using
only the 16 parameters in the table that are indicated by boxes.  To further
simplify the problem, we define the mean surface temperatures as the average
temperature over that surface. For example, $T_{mci}$ is the average
temperature over the inner surface of the inner boundary of the PCM container
(diameter $D_{ci}$). Next we apply the Buckingham Pi
theorem to determine the minimum number of dimensionless groups given the
dimensional parameters in Table \ref{tab:paralist} and the assumption that
mass, length, time, and temperature are independent dimensions. This leads us
to expect 12 dimensionless parameters.  Given the expected number of groups
and the well-established definitions of many of them, we arrive at the
dimensionless groups in Table \ref{tab:grouplist}.
\begin{sidewaystable}
	\centering
	\renewcommand{\arraystretch}{1.5}
	\caption{Non-Dimensional groups affecting heat transfer rates in LHTES devices. The independent groups corresponding to parameters in \ref{tab:paralist} have been boxed. Other dependent groups like the Rayleigh number have been mentioned due to their importance in literature.}
	\begin{tabular}{>{\raggedright\arraybackslash}p{0.15\textwidth}cccc}
		Numbers & HTF   & PCM   & HTF Tube & PCM Container \\
		\hline
		Reynolds number ($Re$) & \fbox{${\rho_f D_t u_f}/{\mu_f}$} & - & - & - \\
		
		Fourier number ($Fo$)	&   - 	&	\fbox{ $\alpha_p t/(D_{co}-D_{ci})^2$ } 	 &	-	&	-\\
		
		Prandtl number ($Pr$) & \fbox{${Cp_f\mu_f}/{k_f}$} & \fbox{${Cp_p\mu_p}/{k_p}$} & - & - \\
		
		P\'eclet number ($Pe$) & ${D_t u_f}/{\alpha_f}$ & - & - & - \\
		
		Grashof number ($Gr$) & - & \fbox{$g\beta_p({D_{co} - D_{ci}})^{3} (T_{mci}-T_{mean})/{\nu_p^2}$} & - & - \\
		
		Rayleigh number ($Ra$) & - & $g\beta_p({D_{co} - D_{ci}})^{3} (T_{mci}-T_{mean})/{\nu_p \alpha_p}$ & - & - \\
		
		Aspect ratio ($AR$) & - & - & \fbox{$l_t/D_t$} & \fbox{$l_c/(D_{co}-D_{ci})$} \\
		
		Stefan number ($Ste$) & - & \fbox{$Cp_p(T_{mci}-T_{mean})/L$} & - & - \\
		Biot number ($Bi$) & - & - & \fbox{$h_f(D_{ci}-D_t)/k_t$} & - \\
		Nusselt number ($Nu$) & \fbox{$h_fD_t/k_f$} & \fbox{$h_pD_{ci}/k_p$} & - & - \\
		Melting to Heat Transfer Timescale & - & \fbox{$\Delta t_s/t \approxeq 0$} & - & - \\
		
	\end{tabular}%
	\label{tab:grouplist}%
\end{sidewaystable}%

\subsection{Relationship between heat transfer rate and melt fraction}
In the foregoing analysis we have sought the minimum number of dimensionless
groups while recognizing that there are multiple ways to define these groups.
In particular, it is useful to consider that the rate of change of the melt
fraction is related by conservation of energy to the heat transfer rate with
the assumption of isothermal heat transfer. The HTF transports energy into the system, which is then transferred to other system components. Let 
the total heat transfer rate to the system be denoted by $q_f$, as given in 
Table \ref{tab:other_paralist}, which is equal to the heat transfer out of the control volume HTF. 
This heat is then distributed between the PCM($q_p$), HTF tube($q_t$) and PCM 
container($q_c$). There will be a transient as the temperatures of the PCM and
heat transfer surfaces adjust to the melting point of the PCM.
Once this transient is finished, $q_t$ and $q_c$ are expected to be small compared to $q_p$ due to the high 
volume and the high heat capacity of the PCM. \qp can be further split into two components, the sensible heating rate \qps which 
causes temperature rise in the PCM and the latent heating rate \qpl which causes 
melting of the PCM. \qps is typically much smaller than \qpl due to reasons similar as 
above; the latent heat capacity of the PCM $L$ is a couple of orders of magnitude 
higher than the sensible heat capacity $Cp_p$. Assuming that the PCM container is well insulated, the heat transfer rate balance can be 
written as
\begin{align}
	q_f &= q_p + q_t + q_c \nonumber \\
		&= \qpl + \qps + q_t + q_c
	\label{eq:htrate}
\end{align}
Out of these, \qpl is of particular interest as it represents high quality energy available at a 
fixed temperature. The integral of \qpl from the onset of melting to the current time 
is related to the melted fraction of PCM ($\eta$) by \eqref{eq:hrate_eta}, where 
$\eta$ is defined as the mass of melted PCM to the total mass of PCM.
\begin{equation}
\label{eq:hrate_eta}
\qpl = M_p L \frac{d\eta}{dt}
\end{equation}
\qpl or $\eta$ can be written as a function of all the parameters in Table
\ref{tab:paralist}.  Depending on how many of those we vary for our
simulations, we get a corresponding number of dimensionless numbers. This is
discussed further in \S\ref{sec:Simulations}.

\section{Numerical simulations approach}
\label{sec:numerical}
To investigate the relationships between the dimensionless groups described in
Table \ref{tab:grouplist}, we seek benchmark simulations that are free from
models.  In practice, some modeling is inherent in simulations, starting with
the continuum approximation, which omits molecular effects inherent in the
phase change process.  Our approach is to limit the modeling in the
simulations to:
\begin{inparaenum}
	\item The HTF is incompressible and Newtonian.
	\item The initial temperature of the entire unit is uniform and the PCM is in the 
	solid phase
	\item The thermophysical properties of the liquid HTF, the PCM and the container are 
	constant except for the density of the PCM.
	\item The density changes in the PCM and their scaling height are
          small so that the
          non-hydrostatic Boussinesq approximation is applicable.
	\item The kinetic and thermal energies of the PCM are decoupled.
        \item The equations of motion for the liquid and solid phases of the
          PCM are coupled using the approach of \citet{voller87a}.
\end{inparaenum}
To make the simulations more tractable, only laminar flow of the HTF and PCM
are considered so that the axisymmetric equations of motion are applicable.

The PCM flow is assumed to satisfy the non-hydrostatic Boussinesq assumptions for
conservation of mass and momentum which can be written in cylindrical
coordinates as
\begin{subequations}
	\label{eq:newtonian_navier_stokes_longeon13}
	\begin{equation}
	\frac { 1 } { r } \frac { \partial \left(r u _ { r } \right) } { \partial r } + 
	\frac { \partial \left(u _ { z } \right) } { \partial z } = 0
	\end{equation}
	
	\begin{equation}
	\begin{aligned} \rho_0 \left( \frac { \partial u _ { r } } { \partial t } + u _ { 
		r } \frac { \partial u _ { r } } { \partial r } + u _ { z } \frac { \partial u _ { 
			r } } { \partial z } \right) = - \frac { \partial p^* } { \partial r } + 2\mu 
	\frac { \partial^2 u _ { r } } { \partial r^2 } \\ + \mu \frac { \partial } { 
		\partial z } \left( \frac { \partial u _ { r } } { \partial z } + \frac { \partial 
		u _ { z } } { \partial r } \right) \\ + \frac { 2 \mu } { r } \left( \frac { 
		\partial u _ { r } } { \partial r } - \frac { u _ { r } } { r } \right) + S_r 
	\end{aligned}
	\end{equation}
	
	\begin{equation}
	\begin{aligned} \rho_0 \left( \frac { \partial u _ { z } } { \partial t } + u _ { 
		r } \frac { \partial u _ { z } } { \partial r } + u _ { z } \frac { \partial u _ { 
			z } } { \partial z } \right)  = - \frac { \partial p^* } { \partial z } + 2\mu 
	\frac { \partial^2 u _ { z } } { \partial z^2 } \\ + \mu \frac { \partial } { 
		\partial r } \left( \frac { \partial u _ { z } } { \partial r }  + \frac { 
		\partial u _ { r } } { \partial z } \right) + \frac { \mu } { r } \left( \frac { 
		\partial u _ { r } } { \partial z } + \frac { \partial u _ { z } } { \partial r } 
	\right) + S_z + S_b\end{aligned}
	\end{equation}
\end{subequations}
Here, the force terms $S_r$, $S_z$ are the momentum sinks used by the 
melting/solidification model of \citet{voller87a} and are given as
\begin{equation}
S_r = A_{mush}\frac{(1-\lambda)^2}{(\lambda^3+\epsilon)}u_r,
S_z = A_{mush}\frac{(1-\lambda)^2}{(\lambda^3+\epsilon)}u_z
  \label{eq:meltingModel}
\end{equation}
where $A_{mush}$ is the mushy zone constant of \citet{voller87a}. In their model, the 
liquid fraction $\lambda$ is calculated as 
\begin{equation}
\lambda = \frac{T_{liquidus}-T}{T_{liquidus}-T_{solidus}} \ .
\end{equation} 
The source term $S_b$ is the buoyancy force given by $\rho_0 \beta (T-T_0)g$, where $T_0$ and $\rho_0$ are the 
reference temperature and reference density used for the Boussinesq approximation and $\beta$ is the 
coefficient of thermal expansion. Within the Boussinesq approximation, viscous heating 
of the fluid is taken to be negligible and so the thermal and mechanical energy
equations decouple. The mechanical energy equation can be derived by taking
the dot product of velocity and momentum.  The thermal energy equation can be
written in terms of enthalpy or temperature. 
ANSYS Fluent, which is the code used for simulations, uses the enthalpy form of the equation, given as 

\begin{equation}
\rho_0 \left( \frac { \partial h } { \partial t } + u _ { r } \frac { \partial h } 
{ \partial r } + u _ { z } \frac { \partial h } { \partial z } \right) = \rho \dot 
{ q } _ { g } + \frac { 1 } { r } \frac { \partial } { \partial r } \left( k r 
\frac { \partial T } { \partial r } \right) \\ + \frac { \partial } { \partial z } 
\left( k \frac { \partial T } { \partial z } \right) + S_e 
\end{equation}
where 
\begin{equation}
S_e = \frac{\partial \rho \Delta H}{\partial t} + \nabla \cdot (\rho \vec{u}\Delta 
H) \ .
\end{equation}
Here 
\begin{equation}
\Delta H = L\lambda 
\end{equation}
is the latent heat enthalpy change for a material volume of PCM.

The work associated with the momentum terms $S_r$ + $S_z$ is
\begin{equation}
  S_e' = \vec{u}\cdot (S_r \hat{r} + S_z \hat{z})
\end{equation}
Due to the small velocities, $S_e' \ll S_e$ and is neglected.

\section{Simulations}
\label{sec:Simulations}
\begin{figure}[h]
	\centering
	\begin{subfigure}[t]{0.4\textwidth}
		\includegraphics[scale=0.7]{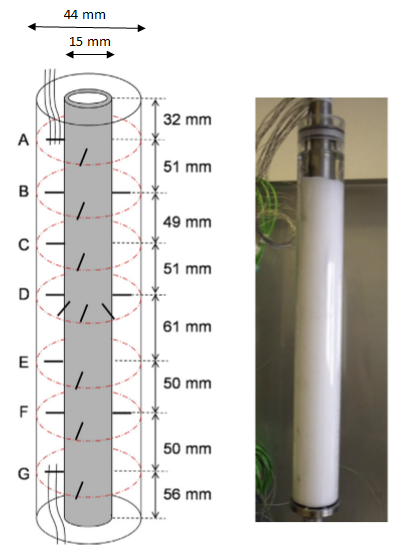}
		\caption{{\footnotesize Experimental setup by \citet{longeon13}}}
		\label{fig:longeon_geom}
	\end{subfigure}
	\begin{subfigure}[t]{0.4\textwidth}
		\includegraphics[scale=0.5]{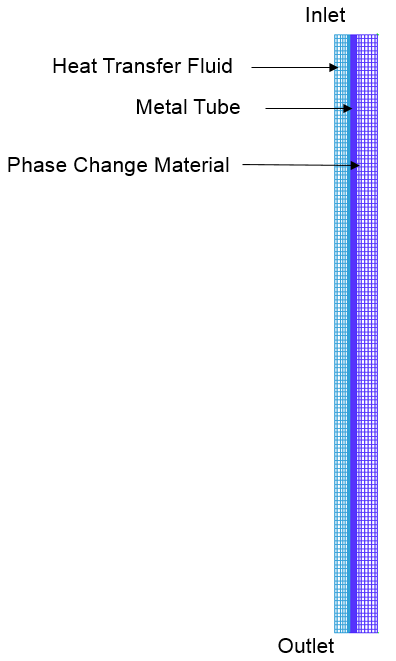}
		\caption{\footnotesize Our computational domain and grid.}
		\label{fig:longeon_compdomain}
	\end{subfigure}
	\caption{Simulation geometry based on the laboratory experiments of \citet{longeon13}. Panel (a) shows their experimental setup, reprinted with their permission.}
\end{figure}

The simulation geometry is chosen for validation against the laboratory
results of \citet{longeon13}.  The physical configuration is shown in Figure
\ref{fig:longeon_geom}.  Due to the fact that the cylinder is oriented
vertically and that the flow regime is laminar, it can be assumed that the
flow is axially symmetric and the equations of motion in \S\ref{sec:numerical}
are applicable.
Details of the computational geometry are in Table
\ref{table:LongeonGeomP}. The structured grid used for the simulations is
shown as Figure 
\ref{fig:longeon_compdomain}. The properties of the PCM, given in Table
\ref{table:longeonpcm}, 
are matched to those in 
\citet{longeon13}, with the exception of 
the sensible specific heat capacity and the density, which are different between
the solid to liquid in the experiments but in the simulations are set to
average values shown in Table \ref{table:longeonpcmsim}. 
The properties of stainless steel in the simulations are
density $8030\ kg/m^3$, specific heat capacity $502.48\ J/kg K$ and thermal 
conductivity of $16.27\ W/mK$. 

\begin{table}[h]
	\centering	
	\begin{subtable}[t]{0.45\linewidth}
		\centering
		\caption{Geometry of simulation domain for validation}
		\label{table:LongeonGeomP}
		\begin{tabular}{lll}
			Parameter 		& Value & Unit\\
			\hline 
			\multicolumn{3}{l}{HTF tube}\\ 
			Outer Radius 	& 10 	& mm\\
			Inner Radius 	& 7.5 	& mm\\
			Length			& 400 	& mm\\
			\hline
			\multicolumn{3}{l}{PCM container}\\ 
			Inner Radius 	& 22 	& mm\\
			Length 			& 400 	& mm
		\end{tabular}
	\end{subtable}
	\begin{subtable}[t]{0.45\linewidth}
		\centering
		\caption{Properties of PCM used for simulation}
		\label{table:longeonpcmsim}
		\begin{tabular}{lll}
			Property    & Value    & Unit   \\
			\hline
			$\rho$        & 820      & $kg/m^3$   	\\
			$L$           & 157      & $kJ/kg$  	\\
			$C_p$          & 2.1      & $kJ/kg.K$  	\\
			$\mu$          & 0.002706 & $kg/m.s$ 	\\
			$\beta$        & 0.001    & $1/K$		\\
			$T_{solidus}$  & 34.95    & $\degree C$	\\
			$T_{liquidus}$ & 35       & $\degree C$	\\
			$k$   		   & 0.2     & $W/m.K$ 
		\end{tabular}
	\end{subtable}
	\\	
	.\\
	\begin{subtable}[t]{0.45\linewidth}
		\centering
		\caption{Properties of PCM RT35 Rubitherm as reported by \cite{longeon13}}
		\label{table:longeonpcm}
		\begin{tabular}{lll}
			Property    & Value    & Unit   \\
			\hline
			$\rho$        & 880(s)/760(l)      & $kg/m^3$   	\\
			$L$           & 157      & $kJ/kg$  	\\
			$C_p$          & 1.8(s)/2.4(l)      & $kJ/kg.K$  	\\
			$\mu$          & 0.002706 & $kg/m.s$ 	\\
			$\beta$        & 0.001    & $1/K$		\\
			$T_{m}$ 		& 35       & $\degree C$	\\
			$k$   		   & 0.2     & $W/m.K$ 
		\end{tabular}
	\end{subtable}
	\begin{subtable}[t]{0.45\linewidth}
		\centering
		\caption{Properties of HTF Water}
		\label{table:longeonhtf}
		\begin{tabular}{lllll}
			Property    & Value    & Unit	\\
			\hline
			$\rho$ 	& 998.2     & $kg/m^3$	\\
			$C_p$   & 4.182     & $kJ/kg.K$	\\
			$\mu$   & 0.001003 	& $kg/m.s$	\\
			$k$     & 0.6      	& $W/m.K$	\\
		\end{tabular}
	\end{subtable}
	\caption{Simulation parameters used in \citet{longeon13}}
\end{table}

\subsection{Numerical details}
The simulations are conducted using the finite volume code ANSYS Fluent. The
simulation parameters are in Tables \ref{table:longeonpcmsim} and \ref{table:longeonhtf}.
In the phase-change model \eqref{eq:meltingModel},
the constant $A_{mush}$ defining the mushy zone is taken to be 100,000 and it is
observed that the solution is not strongly dependent on this
value. 

The HTF inlet boundary condition is defined to be a uniform velocity of
$0.01\ m/s$ and with static temperature 53$ \degree$C.  The HTF outlet
boundary condition is constant gauge pressure of $0\ Pa$.  The internal walls
of the tube and PCM container are conjugate heat transfer internal boundaries with no slip. Heat transfer between the outer walls of PCM container and the room is ignored due to the low
temperature differences between the heat transfer medium and ambient
conditions. Thus, the outer walls are adiabatic with no slip. The vertical
axis of the heat transfer fluid tube is defined to be a symmetry boundary
condition so that the simulations are axisymmetric.

The mass and momentum equations are solved using the pressure based solver with the
SIMPLE algorithm used for the pressure velocity coupling. Pressure is
discretized using the PRESTO scheme \citep{shmueli10}. The momentum and energy equations 
are discretized using second order upwind schemes. The evolution in time is 
first order implicit, as it is sufficient for most problems \citep{ANSYSFluent2011}. 
The solution is initialized with zero velocity in all directions and an ambient 
temperature of 23$\degree$C. The highest velocity in the domain is expected in the 
HTF and is twice the mean velocity, for a fully 
developed flow, which is $0.02\ m/s$.

\subsection{Validation simulations}
\begin{figure}
	\centering
	\includegraphics[width=\textwidth]{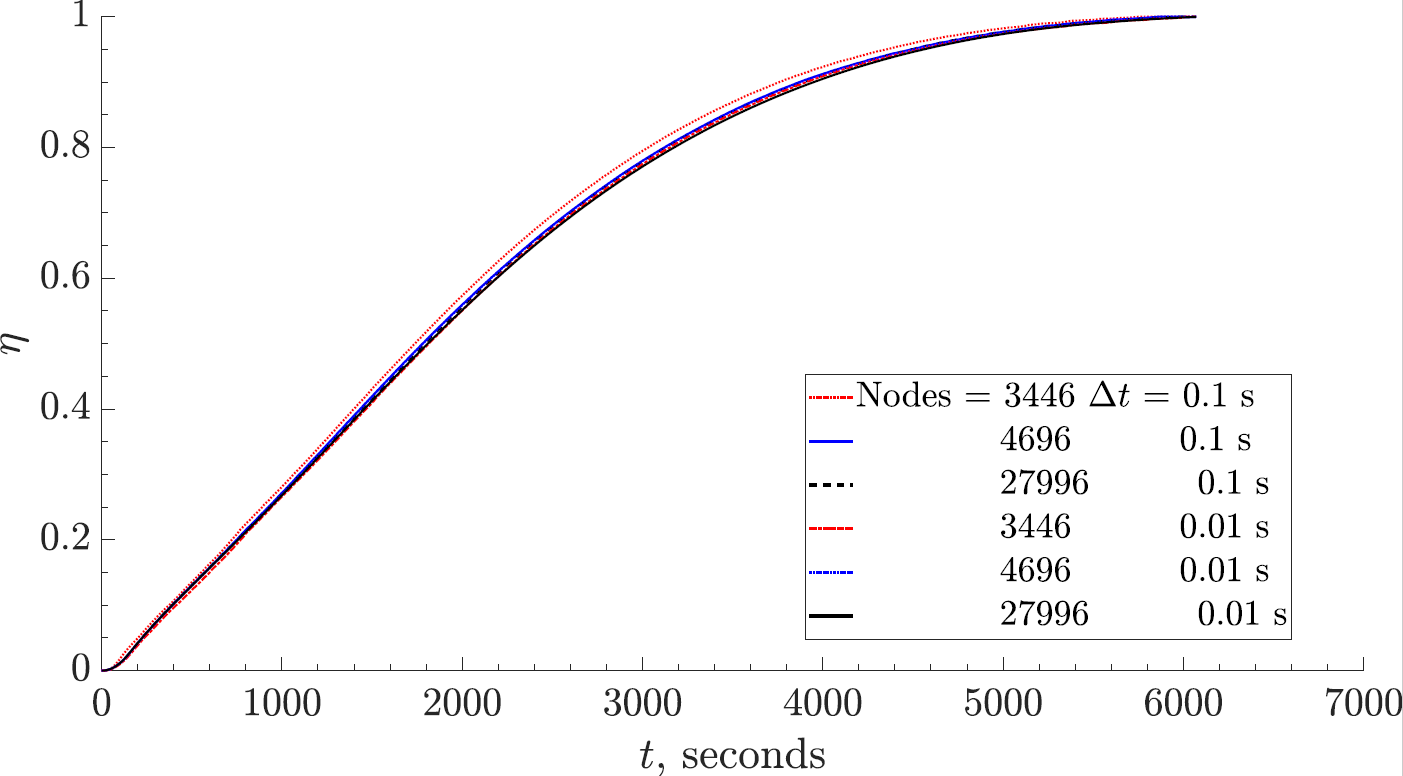}
	\caption{Grid Sensitivity, Average Melt Fraction in PCM}
	\label{fig:gridsenmelt}
\end{figure}

Sensitivity of the solutions to grid resolution and time step size are examined by varying the time step size by two orders of magnitude and the
number of finite volumes in the grid by a factor of approximately eight. The important variable for energy storage is the melted fraction of the PCM, given as $\eta$. We perform simulations with grid sizes $3446$, $4696$ and $27996$ and time-step sizes of $0.1$ and $0.01$ seconds. Note that our smallest grid size and largest time-step are the same order of magnitude as the grid size of $9000$ and time-step of $0.5$ seconds used by \citet{longeon13}. Figure
\ref{fig:gridsenmelt} shows the melt fraction for the three grid sizes and two different time step ($\Delta t$) sizes. The results show that with increasing spatial and temporal resolution, the curves approach the results for the finest grid of $27996$ and the finest time-step size of $0.01$ seconds, and the difference between the intermediate resolution of $4696$ and $0.01$ seconds and the finest resolution is negligible. Table \ref{table:gridindepmelt} shows the maximum of the absolute error(as percentage) in $\eta$ in reference to the finest resolution case, and we see that the error reduces by less than 1\% beyond the intermediate resolution of $4696$ and $0.01$ seconds. Thus, the grid size of 
$4696$ nodes and a time step of $\Delta t=0.01$ seconds is sufficient to obtain grid 
insensitive results. 

\begin{figure}
	\centering
	\includegraphics[width=\textwidth]{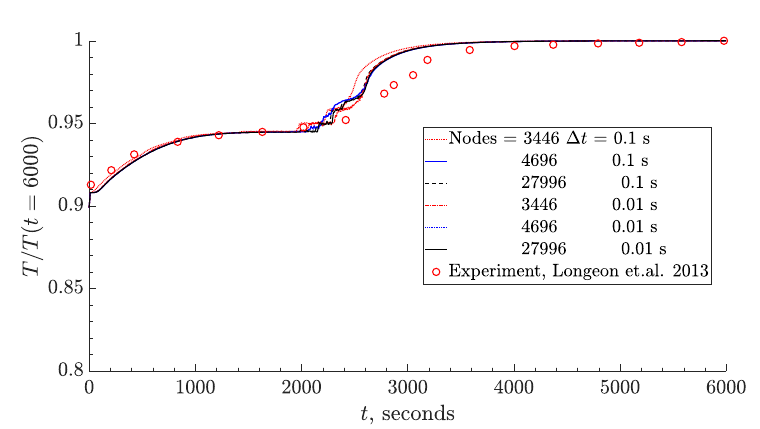}
	\caption{Comparison of measured temperature at Point D\citep{longeon13} with 
simulation for validating simulation procedure}
	\label{fig:valtemppointD}
\end{figure}

Figure \ref{fig:valtemppointD} shows the comparison of temperature at a specific 
location D obtained from simulations and measured experimentally by \citet{longeon13}. We see good agreement between the shapes of the 
experimental and numerical data curves.

\begin{table}
	\centering
	\caption{Grid sensitivity for melt fraction $\eta$, difference relative to nodes$=27996$ and $\Delta t = 0.01$}
	\begin{tabular}{ll|ll}
		$\eta$                               & \multicolumn{3}{c}{Time Step (s)} \\
		\multirow{4}{*}{Grid Size (nodes, cells)}  &        & 0.1     & 0.01   \\
		\hline
		& 3446, 3170   & 2\%  & 1\% \\
		& 4696, 4356   & 1\% & 1\% \\
		& 27996, 26733  & 1\% & 0     
	\end{tabular}
	\label{table:gridindepmelt}
\end{table}

\subsection{Parametric study}
Given the results of the validation experiments, we conclude that the
simulation technique is adequate.
After validating our simulation procedure, we proceeded to our parametric study.  We vary the parameters $u_f$, $T_{in}$, $k_p$ and 
$k_f$ for the geometry and setup discussed in \citet{longeon13}, which we use for 
validating our simulations described in \S\ref{sec:Simulations}. $u_f$ is flow velocity 
for HTF and is the easiest to change through the use of a pump. $T_{in}$ is the inlet 
temperature of the HTF, which depends on the system from which we are extracting 
energy. $k_p$ and $k_f$ are dependent on material properties and additive enhancements 
and we have moderate control over them. Table \ref{tab:caselist} shows the endpoint 
values of parameters changed and table \ref{tab:cases} shows the corresponding nominal dimensionless numbers.
\begin{table}
	\centering
	\caption{Physical parameters corresponding to values of dimensionless numbers in table \ref{tab:cases}. There are a total of 64 simulations created by varying each number in table \ref{tab:cases} independently. For reasons of space, only the endpoint cases (numbering 16) have been shown here.}
	\label{tab:caselist}
	{\small \begin{tabular}{rrrrrrrr}
\multicolumn{1}{c}{$u_f$} & \multicolumn{1}{c}{$k_f$} & \multicolumn{1}{c}{$k_p$} & \multicolumn{1}{c}{$T_{in}$} & \multicolumn{1}{c}{$Re_f$} & \multicolumn{1}{c}{$Pr_f$} & \multicolumn{1}{c}{$Pr_p$} & \multicolumn{1}{c}{$Gr_p$} \\
\hline

  0.01 & 0.1 & 0.1 & 310.125 & 149 & 42 & 57 & 24906  \\
  0.01 & 0.1 & 0.1 & 324.125 & 149 & 42 & 57 & 199248  \\
  0.01 & 0.1 & 1 & 310.125 & 149 & 42 & 6 & 24906  \\
  0.01 & 0.1 & 1 & 324.125 & 149 & 42 & 6 & 199248  \\
  0.01 & 0.8 & 0.1 & 310.125 & 149 & 5 & 57 & 24906  \\
  0.01 & 0.8 & 0.1 & 324.125 & 149 & 5 & 57 & 199248  \\
  0.01 & 0.8 & 1 & 310.125 & 149 & 5 & 6 & 24906  \\
  0.01 & 0.8 & 1 & 324.125 & 149 & 5 & 6 & 199248  \\
  0.14 & 0.1 & 0.1 & 310.125 & 2090 & 42 & 57 & 24906  \\
  0.14 & 0.1 & 0.1 & 324.125 & 2090 & 42 & 57 & 199248  \\
  0.14 & 0.1 & 1 & 310.125 & 2090 & 42 & 6 & 24906  \\
  0.14 & 0.1 & 1 & 324.125 & 2090 & 42 & 6 & 199248  \\
  0.14 & 0.8 & 0.1 & 310.125 & 2090 & 5 & 57 & 24906  \\
  0.14 & 0.8 & 0.1 & 324.125 & 2090 & 5 & 57 & 199248  \\
  0.14 & 0.8 & 1 & 310.125 & 2090 & 5 & 6 & 24906  \\
  0.14 & 0.8 & 1 & 324.125 & 2090 & 5 & 6 & 199248  \\
\end{tabular}}
\end{table}%
\begin{table}
	\centering
	\caption{List of dimensionless numbers in parameter space and their values under study. Since the parameter space is four-dimensional, the total number of simulations are 64.}
	\label{tab:cases}
	{\small \begin{tabular}{rrrrrrrr}
\multicolumn{2}{c}{$Re_f$} & \multicolumn{2}{c}{$Pr_f$} & \multicolumn{2}{c}{$Pr_p$} & \multicolumn{2}{c}{$Gr_p$} \\
\hline

 $Re_f1$ & 149 &$Pr_f1$ & 42 &$Pr_p1$ & 57 &$Gr_p1$ & 24906  \\
 $Re_f2$ & 2090 &$Pr_f2$ & 5 &$Pr_p2$ & 14 &$Gr_p2$ & 83020  \\
 - & - &- & - &$Pr_p3$ & 8 &$Gr_p3$ & 141134  \\
 - & - &- & - &$Pr_p4$ & 6 &$Gr_p4$ & 199248  \\
\end{tabular}}
\end{table}%

For example, for the four parameter case described here,
\begin{equation}
\label{eq:fc}
\qpl = \qpl(u_f,k_f,k_p,T_{in})
\end{equation}
The variable of interest is the stored energy, which is given by the time integral of 
\qpl. Its dimensionless equivalent is $\eta$, the melt fraction, which we shall use 
henceforth for presenting results.

\section{The structure of the melt fraction curve $\eta(t)$}
\label{sec:regions}
In this section, we discuss some observations
about the structure of $\eta$ prior to looking at the scaled results in section \ref{sec:Results}. We observed two distinct regions for $\eta$, specifically a
linear and an asymptotic region, and name the melt fraction at the transition
between these regions as the {\em transition melt fraction} denoted by
\etac.  The time at which $\etac$ occurs is denoted by $\tau_{critical}$; see section \ref{sec:Results} for further discussion about obtaining a dimensionless time $\tau$ from $t$.  We elaborate in \S\ref{sec:lin_behave} and \S\ref{sec:exp_behave} on
why each region can be expected from physical reasoning.  

\subsection{Linear region}
\label{sec:lin_behave}
Based on Newton's law of cooling, the heat transfer rate $q_f$ is set by the
heat transfer coefficient $h_f$ and the temperature difference
$T_{in}-T_{mci}$. Between the onset of melting and the transition melt
fraction \etac, the temperature $T_{mci}$ is roughly constant due provided
that transport of heat by the HTF does not limit the heat transfer into the
PCM.  If the variation is not significant compared to the total temperature
difference $T_{in}-T_{mci}$, $T_{mci}$ and the difference can be considered to
be constant. Indeed, for constant wall temperature with internal laminar flow,
the non-dimensional heat transfer coefficient, which is the Nusselt number, is
constant with value equal to 3.66 \citep[eq. 8.55]{incropera11}.  Thus, $q_f$ is
expected to be a constant in this temperature region.

In the PCM, \qp, which is comparable to $q_f$, is proportional to the
temperature difference $T_{mci} - T_{mean}$, which is also constant. After an
initial transient, \qpl is the major component of \qp. The melt fraction
$\eta$, which is proportional to integral of \qpl as shown in
\eqref{eq:hrate_eta}, is expected be linear with time.  
At \etac, the quantity of solid PCM gets small such that the
characteristic temperature difference in the PCM is $T_{mci} - T_p$, where
$T_p$ is the mean temperature of the PCM and is approximately equal to the far
field temperature. $T_p$ is rising inverse-exponentially, which results in the
asymptotic behavior of $\eta$, as explained in the following section.
\subsection{Exponential region}
\label{sec:exp_behave}
\begin{figure}
	\centering
	\includegraphics[width=0.5\textwidth]{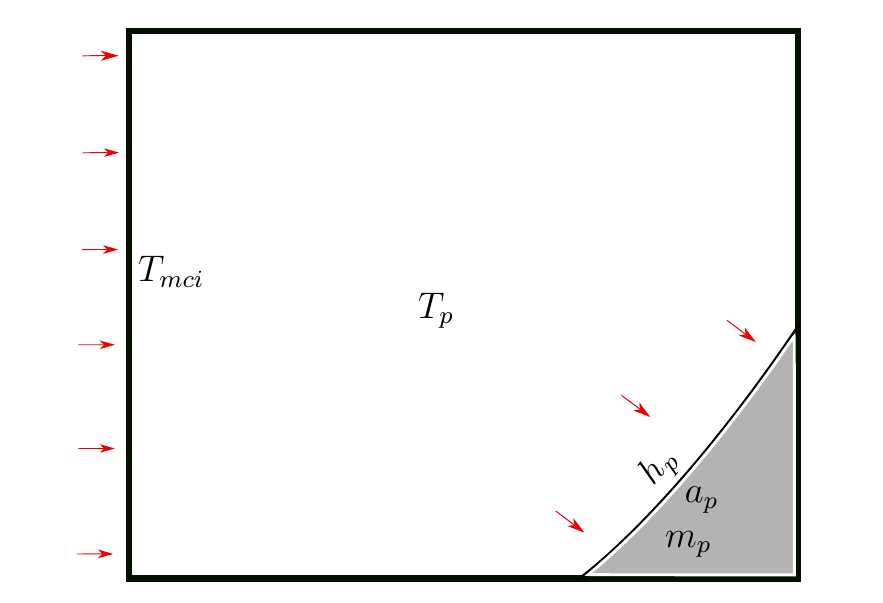}
	\caption{Cartoon figure showing the variables for the asymptotic model. The interface between $m_{pcm}$ and the liquid is an arbitrarily drawn curve.}
	\label{fig:cartoon_exp_eta}
\end{figure}

Since the purpose of this section is to analze behavior rather than predicting data 
from first principles, we shall use simplified notation to obtain uncluttered 
equations. Terms expected to be constant have been grouped into numbered constants for 
brevity. Let the mass of solid PCM at the time $\eta$ reaches \etac be $m_{pcm}$, and 
let its surface area be $a_{pcm}$. Let the heat transfer coefficient on the solid 
liquid interface be $h_{pcm}$. Figure \ref{fig:cartoon_exp_eta} shows a cartoon 
representation of the variables of interest. 

After $\eta$ reaches a critical fraction \etac, the mass of solid PCM is small and the 
characteristic temperature difference is closer to $T_{mci} - T_p$ rather than 
$T_{mci} - T_{mean}$.  The configuraton is shown in cartoon form in
Figure \ref{fig:cartoon_exp_eta} along with the notation used in the following discussion
In this regime, there is limited contact area between the liquid and the solid
and so most of the heat transferred from the HTF to the PCM raises the
temperature of the PCM.  With this approximation, 
\begin{equation}
	\begin{aligned}
	\qp = M_p Cp_p \frac{dT_p}{dt} &= h_p A_p (T_{mci}-T_p)  \\
	\Rightarrow \frac{d(T_{mci}-T_p)}{dt} &= -\frac{h_p A_p}{M_p Cp_p} (T_{mci}-T_p) \\
	\Rightarrow \frac{d(T_{mci}-T_p)}{T_{mci}-T_p} &= -c_2' dt \\
	\Rightarrow ln(T_{mci}-T_p) &= -c_2't + c_1 \\
	\Rightarrow T_p &= T_{mci} - c_2 e^{(-c_2't)}
	\end{aligned}
	\label{eq:exp_T_p}
\end{equation}

In short, the PCM acts as a lumped capacitance \citet[eq. 5.8a]{incropera11}
because the 
mass of the solid PCM is insufficient to affect $T_p$. Since the purpose
of the foregoing is to arrive at the expected functional form rather than a
numerically exact model, we have combined terms that are approximately
constant into the coefficients the coefficients $c_1$, $c_2$, $c_2'$. The prime notation denotes constants that carry forward into the final expression given in \eqref{eq:exp_eta}. 

Assuming that the remaining solid PCM is at melting temperature and there is no 
significant sensible heating of the residual solid, the heat transfer to the
solid PCM is
\begin{equation}
	\begin{aligned}
	\qpl = L\frac{d m_{pcm}}{dt} &= h_{pcm} A_{pcm} (T_p - T_{mean})\\ 
	\Rightarrow \frac{d(1-\eta)}{dt} &= \frac{h_{pcm} A_{pcm}}{LM_p} (T_p
        - T_{mean}) \ .
	\end{aligned} 
	\label{eq:exp_eta_1}
\end{equation}
The area $A_{pcm}$ depends on the mass of solid PCM and can be calculated if
the the shape of $m_{pcm}$ and its density is known. Since the PCM is close to
the melting temperature, the density can be considered to be a constant. At
constant density, if $m_{pcm}$ is a sphere, $A_{pcm}$ is proportional to
$m_{pcm}^{2/3}$. In general, $A_{pcm}$ is proportional to $m_{pcm}^\gamma$
where $\gamma$ is some real number less than 1, expected to be constant if
the melting front geometry and density do not change in the duration of the
exponential melting region. Substituting this and \eqref{eq:exp_T_p} into
\eqref{eq:exp_eta_1} yields
\begin{equation}
	\begin{aligned}
	\frac{d(1-\eta)}{dt} &= \frac{c_3 h_{pcm} m_{pcm}^p}{LM_p} (T_p -
        T_{mean})\\ \Rightarrow \frac{d(1-\eta)}{dt} &= \frac{c_3 h_{pcm}
          (1-\eta)^\gamma}{LM_p^{1-\gamma}} (T_p - T_{mean})\\ \Rightarrow
        \frac{d(1-\eta)}{(1-\eta)^\gamma} &= \frac{c_3 h_{pcm}
        }{LM_p^{1-\gamma}} (T_p - T_{mean}) dt\\ &= c_5 \left((T_{mci} - T_{mean})
        - c_2 e^{(-c_2't)} \right) dt\\ &= c_5(T_{mci} - T_{mean})dt - c_2 c_5
        e^{(-c_2't)} dt\\ \Rightarrow \frac{(1-\eta)^{\gamma+1}}{\gamma+1} &=
        c_5(T_{mci} - T_{mean})t + c_6 e^{(-c_2't)} + c_7\\ \Rightarrow 1-\eta
        &= \left(c_7' + c_5'(T_{mci} - T_{mean})t + c_6'
        e^{(-c_2't)} \right)^{\frac{1}{\gamma+1}}
	\end{aligned} 
	\label{eq:exp_eta_2}
\end{equation}
From this we conclude that the function form for $\eta(t)$ is 
\begin{equation}
	\eta = 1 - \left(c_7' + c_5'(T_{mci} - T_{mean})t + c_6' 
e^{(-c_2't)} \right)^{\frac{1}{\gamma+1}}
	\label{eq:exp_eta} \ .
\end{equation}
Again, we are interested in the form of the equation and
the numbered coefficients collect constant terms that would make the form more
difficult to read if included in full.

\begin{figure}
	\centering
	\includegraphics[width=0.7\textwidth]{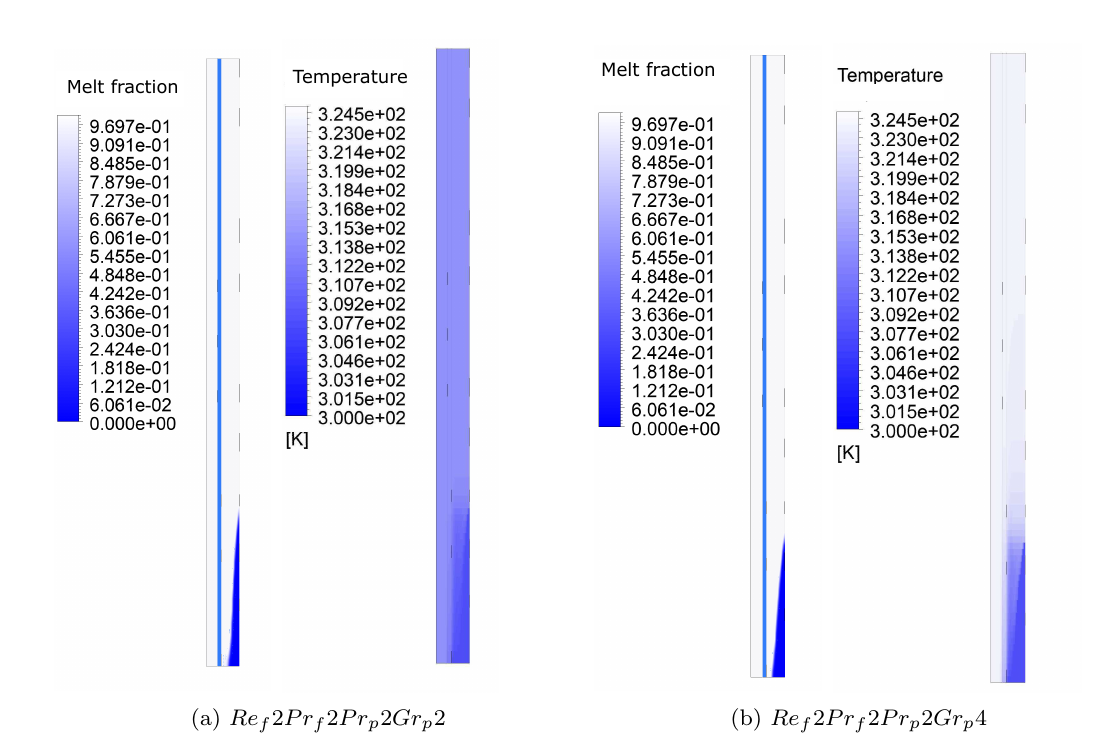}
	\caption{Plot of liquid fraction at time when \etac is reached for two Grashof 
		numbers at $Re_f2$, $Pr_f2$ and $Pr_p2$. The corresponding temperature plot is 
		shown to the right. Notice the similarity between the shape of the solid PCM for 
		the two different cases. The figures are best viewed in conjunction with the geometry and grid shown in figure \ref{fig:longeon_compdomain}.}
	\label{fig:contour_etac1}
\end{figure}

\begin{figure}
	\centering
%
	\includegraphics[width=0.7\textwidth]{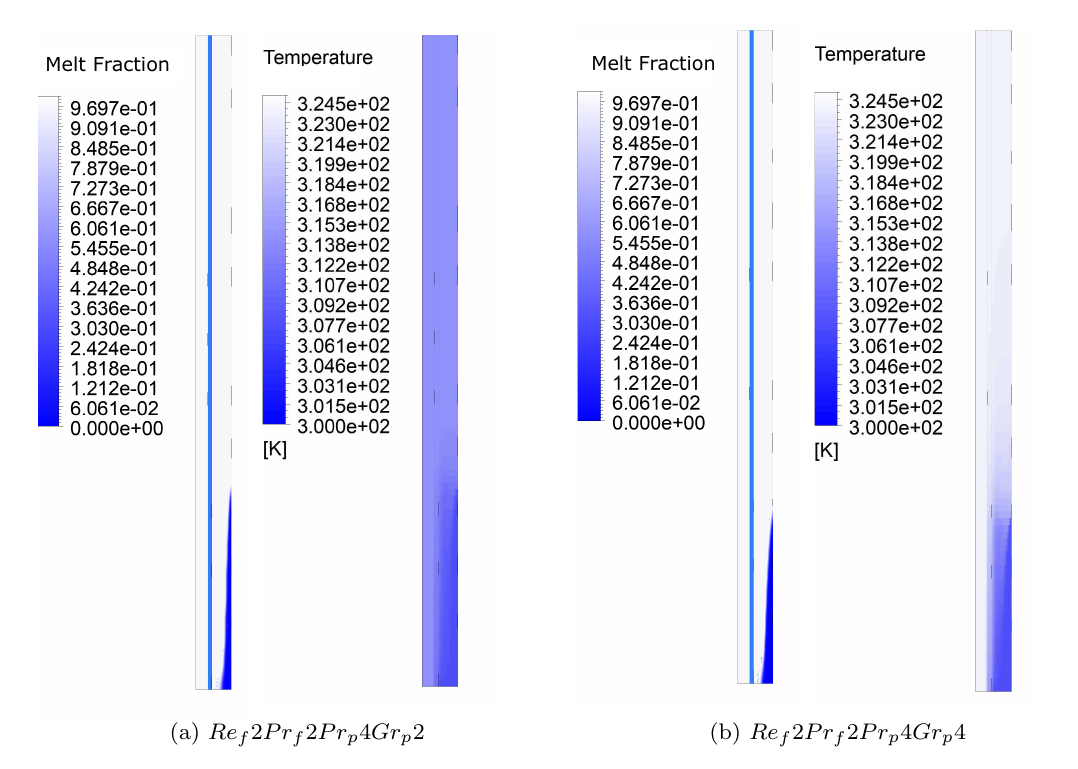}
	\caption{Plot of liquid fraction at time when \etac is reached for two Grashof 
		numbers at $Re_f2$, $Pr_f2$ and $Pr_p4$. The corresponding temperature plot is 
		shown to the right. Notice the similarity between the shape of the solid PCM for 
		the two different cases and the cases from figure \ref{fig:contour_etac1}}
	\label{fig:contour_etac2}
\end{figure}

Figures \ref{fig:contour_etac1} and \ref{fig:contour_etac2} show the melt fraction and temperature contours when melting has reached \etac, for cases with different $Gr_p$ and $Pr_p$ values. The cases have vastly different operating parameters, but we can see that there are similarities in the melt fraction profiles, for example, the shape of the remaining PCM, which has been identified in \S\ref{sec:exp_behave} as a factor in determining the shape of the melt fraction curve. This suggests that \etac might be universal for a given device, at least in the range of dimensionless numbers studied. 

From the arguments in \S\ref{sec:lin_behave} and \S\ref{sec:exp_behave}, we
expect $\eta$ to vary linearly in time when $\eta$ is small and to vary
asymptotically with time when $\eta$ is large with the transition between the
two regimes being \etac.  This conclusion is based entirely on physical
reasoning.  In the following section, we apply this physical reasoning to the simulation data to
understand the time evolution of $\eta$ and, in particular, how this time
scales with the dimensional quantities in Table \ref{tab:paralist}, and to
determine the empirical value of $\etac$, which is of practical importance.

\section{Application of methodology to understand flow physics}
\label{sec:Results}

In the previous sections, we develop our approach for defining the
dimensionless groups important for describing a simple LHTES systems such that
they can be related to the heat transfer rate in a way consistent with the
physical understanding developed from the theory of heat and mass transfer as
well as the significant body of literature on LHTES systems. Here we
demonstrate the utility of the approach by examining $\eta$ as a function of
time in the multi-dimensional space defined $Gr_p$, $Pr_p$, $Re_f$ and
$Pr_f$. The melt fraction $\eta$ has been defined previously as a function of
dimensional time. For consistency, we denote the equivalent of $\eta$ that
accepts a generic dimensionless time $\tau$ as an argument, by
$\eta_{\tau}$. The goal is to find $\eta_{\tau} = \eta_{\tau}(\tau)$ with the
$\tau$ defined in terms of a physically relevant time scale. If we are
successful in this then the curves for $\eta_{\tau}(\tau)$ corresponding to
different values of the parameter being varied will collapse to a single curve.
In this section, we attempt to define a suitable $\tau$ based on the flow physics and observations from literature. As
described in \ref{sec:parametrizing_the_problem}, we use the Buckingham-Pi
theorem to organize our approach. A general dimensionless timescale can be defined as follows - 
\begin{equation}
	\tau = \tau(Fo_p,Gr_p,Pr_p,Re_f,Pr_f)
	\label{eq:generaltimeScale}
\end{equation}
where $Fo_p$ is the PCM Fourier number defined in table \ref{tab:grouplist}. 

\begin{figure}
  \begin{center}
  \includegraphics[width=\textwidth]{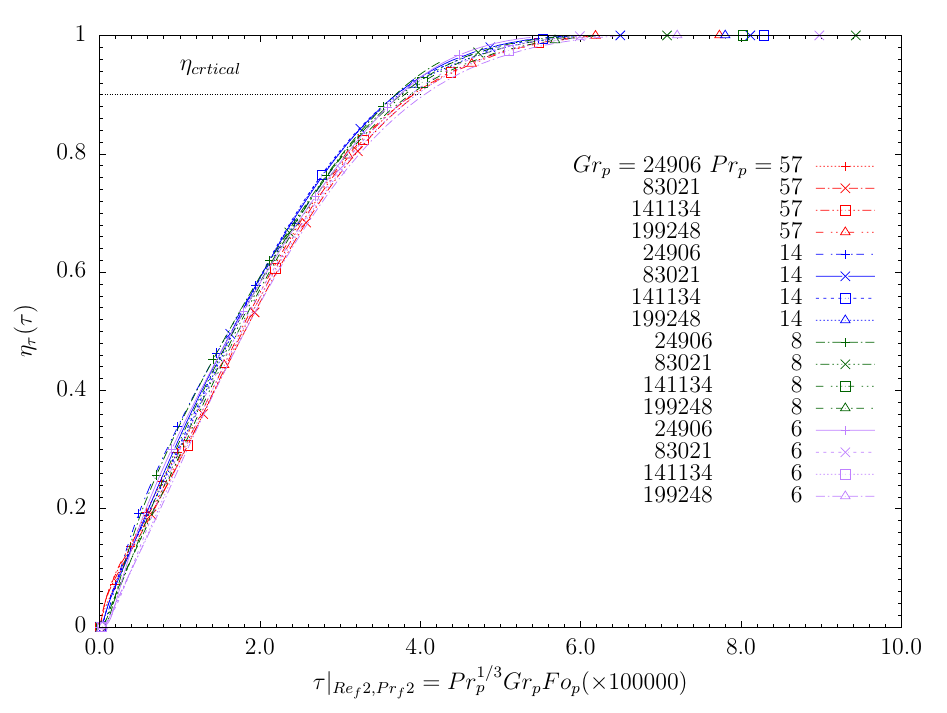}
  \end{center}
  \caption{\label{fig:Grp_Prp} The melt fraction vs dimensionless time on the parameter plane with
    $Re_{f}=2090$ and 
    $Pr_{f}=5$. All curves collapse, indicating that the timescale defined in \eqref{eq:timeScaling} is appropriate if $Re_f$ is high enough and $Pr_f$ is low enough, or if $u_f$ and $k_f$ are both high enough to ensure sufficient heat flow. The markers are plotted to distinguish the curves.}
\end{figure}

The simulation data base consists of 64 cases with parameters tabulated in Table
\ref{tab:cases}.  The simulations span a four-dimensional parameter space
defined by $Re_f$, $Pr_f$, $Pr_p$ and $Gr_p$ with a high and low value for the first two and 4 values each for the rest. As mentioned in \S\ref{sec:lin_behave}, we expect only a weak effect due to $Re_f$ and $Pr_f$, as long as there is sufficient heat flowing into the HTF domain. To confirm this, we perform additional simulations with two more intermediate values of $Re_f=796$ and $Re_f=1443$ at $Pr_f=42$, $Gr_p=8.302\times10^4$ and $Pr_p = 57$, the results of which are shown in figure \ref{fig:Ref}. Similarly, we conduct simulations of two intermediate values of $Pr_f=8$ and $Pr_f=6$ for the case with $Re_f=2090$, $Pr_p=57$ and $Gr_p=8.302\times10^4$, results of which are shown in figure \ref{fig:Prf}. Given the expected and demonstrated weak effect of $Re_f$ and $Pr_f$, we fix their values and apply the
methodology from \S\ref{sec:regions}, beginning with a cut through the
parameter space along the plane defined by $Re_f=2090$ and $Pr_f=5$, that is, a
particular set of HTF parameters.  Based on the reasoning in
\S\ref{sec:regions} along with measurement data from the literature reviewed
in \S\ref{sec:Intro}, we expect time to scale with 
\begin{equation}
  \tau|_{Re_f,Pr_f=constant} = Pr_p^{1/3} Gr_p Fo_p
  \label{eq:timeScaling}
\end{equation}
for fixed $Re_f$ and $Pr_f$ indicated by the subscripts $Re_f,Pr_f=constant$. 
For brevity, we also introduce a shorthand notation for a one dimensional slice through $\tau$, where all parameters except one are kept constant. For example, if all parameters except the PCM Grashof number $Gr_p$ were held constant, $\tau$ would be given as
\begin{equation}
\tau|_{Re_f,Pr_f,Pr_p=constant}(Gr_p) = \taupg
\label{eq:timeScalingshorthand}
\end{equation}
where the subscript $p$ denotes PCM and the additional subscript $g$ denotes that the Grashof number is the variable in question.
To verify the hypothesis of \eqref{eq:timeScaling}, the 
data are plotted with this and other time scalings in Figure
\ref{fig:Grp_Prp}.  In figure \ref{fig:Grp_a}, it is observed that
$Gr_p Fo_p$ collapses the data to a single curve provided that
$Pr_p$ is constant but from figure \ref{fig:Grp_b} it is apparent that the collapse also occurs for multiple values of $Pr_p$. Similarly, figure \ref{fig:Prp} shows that $Pr_p ^{1/3} Fo_p$ collapses the data to a single curve provided that $Gr_p$ is constant. These two relationships are combined in figure \ref{fig:Grp_Prp} and
time is scaled according to
\eqref{eq:timeScaling} to collapse to almost a single curve all 16 cases
having the same values for $Re_f$ and $Pr_f$.

\begin{figure}
	\centering
			\includegraphics[width=\textwidth]{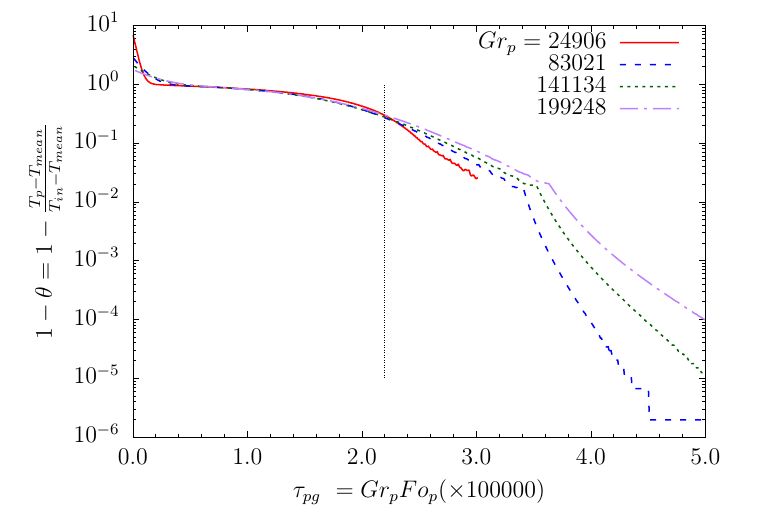}
	\caption{Dimensionless temperature plotted as $1-\theta$ at constant $Re_f=2090$, 
$Pr_f=5$ and $Pr_p=6$. The figure shows the existence of both linear and inverse-exponential regions as described in \S\ref{sec:lin_behave} and \S\ref{sec:exp_behave}. The curves collapse in the linear region. They diverge at the onset of the inverse-exponential behavior, and are displayed as straight lines due to the log-linear axes.}
	\label{fig:Theta_Grp}
\end{figure}

A question that cannot be answered with the approach in
\S\ref{sec:regions} is the value of $\eta_{critical}$. This value is needed in
order to inform whether the linear or asymptotic scaling of $\eta$ with time
is appropriate.  From observing figure \ref{fig:Grp_Prp}, we estimate
$\eta_{critical} \approx 0.9$. The existence of the linear and inverse-exponential regions is further supported by figure \ref{fig:Theta_Grp}, which clearly shows the collapsed curves diverging as straight lines on the onset of the inverse-exponential region, as expected on a log-linear axes. Additionally, figures \ref{fig:Prf}, \ref{fig:Grp_a} and \ref{fig:Grp_b} show that \etatau 
transitions to non-linear around \etac. We have noted in several places in this paper
that the simulations are limited to laminar flow in the PCM whereas practical
systems may employ turbulent flow.  We know the reason, though, why the
value of $\eta_{critical}$ will depend on the hydrodynamic regime of the PCM. In the remaining parts of this section, we discuss the scaling obtained and its practical implications. 

\subsection{Effects of $Gr_p$ and $Pr_p$: Discussion and implications}
\begin{figure}
		\begin{minipage}[t]{0.5\textwidth}
		\includegraphics[width=0.9\textwidth]{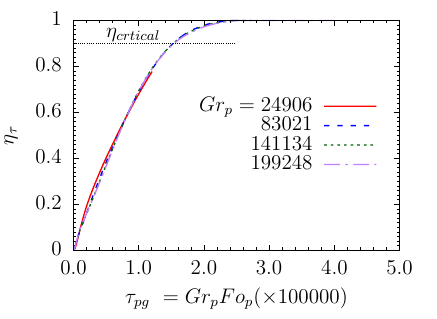}
		\caption{The scaling of melt fraction with the Grashof number on the parameter plane with $Re_{f}=2090$, $Pr_{f}=5$, $Pr_{p}=14$}
		\label{fig:Grp_a}
	\end{minipage}
	\begin{minipage}[t]{0.5\textwidth}
		\includegraphics[width=0.9\textwidth]{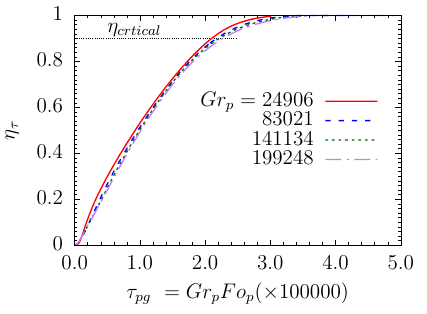}
		\caption{Analogous to figure \ref{fig:Grp_a}, except $Pr_p=6$. }
		\label{fig:Grp_b}
	\end{minipage}
\\
	\begin{minipage}[t]{0.5\textwidth}
				\includegraphics[width=0.9\textwidth]{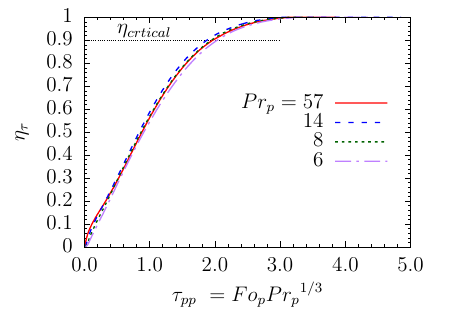}
		\caption{Melt Fraction $\eta$ as a function of $Pr_p$ at constant $Gr_p=199248$, $Pr_f=5$ 
			and $Re_f=2090$. The figure shows the perfect scaling with ${Pr_p}^{1/3}$}
		\label{fig:Prp}
	\end{minipage} %
	\begin{minipage}[t]{0.5\textwidth}
				\includegraphics[width=0.9\textwidth]{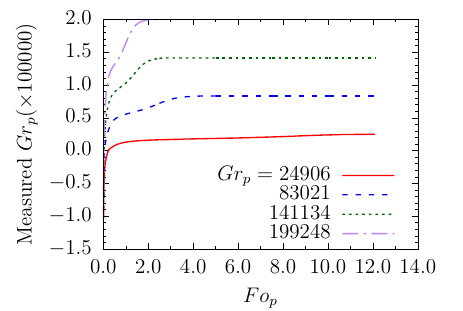}
		\caption{Measured $Gr_p$ as a function of non-dimensional time $\tau$, at constant 
			$Re_f=2090$, $Pr_f=5$ and $Pr_p=6$. The plateau shows that the assumptions from \S\ref{sec:lin_behave} are justified. Negative Grashof numbers indicate that melting temperature has not been reached.}
		\label{fig:Actual_Grp}
	\end{minipage}%
\\
\begin{minipage}[t]{0.5\textwidth}
			\includegraphics[width=0.9\textwidth]{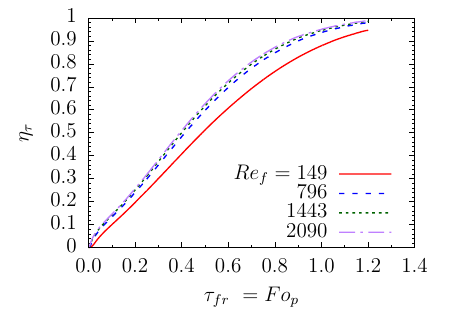}
	\caption{Melt Fraction as a function of $Re_f$ at constant $Gr_p=83020$, $Pr_f=42$ and $Pr_p=57$. The plots show that the heat transfer rate does not improve much past $Re=800$.}
	\label{fig:Ref}
\end{minipage} %
\begin{minipage}[t]{0.5\textwidth}
			\includegraphics[width=0.9\textwidth]{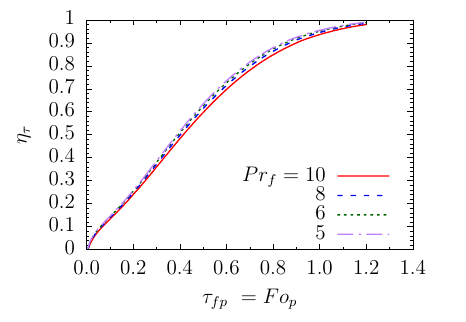}
	\caption{Melt Fraction as a function of four $Pr_f$ at constant $Gr_p=83020$, $Pr_p=57$ and 
		$Re_f=2090$}
	\label{fig:Prf}
\end{minipage}%
\end{figure}

In this section, we discuss the reasons for the scaling obtained in figure \ref{fig:Grp_Prp} and the implications of this scaling for the design and operation of LHTES devices. Figures \ref{fig:Grp_a} and \ref{fig:Grp_b} shows plots for various cases from table \ref{tab:cases} and 
we can see that the different cases collapse to one curve when the time is scaled with 
the PCM Grashof number, which agrees with our predictions in \ref{sec:lin_behave}. 
This is consistent with what \citet{bejan88} observed by scaling analysis for mixed conduction-convection flow 
regimes in an enclosure with laminar flow. On reaching \etac, the curve changes shape 
from linear to asymptotic, as predicted in \ref{sec:exp_behave}. In order to further confirm our hypothesis from \ref{sec:lin_behave}, we plot the measured Grashof number in figure \ref{fig:Actual_Grp}, which indicates that the PCM container wall temperature $T_{mci}$ is indeed constant for the cases under consideration. Figure \ref{fig:Prp} further confirms $Pr_p^{1/3}$ scaling obtained in figure \ref{fig:Grp_Prp}. This scaling corresponds to the $Pr^{1/3}$ scaling observed in laminar forced convection with uniform heat flux. As explained in section \ref{sec:lin_behave}, both the temperature difference $T_{mci}-T_{mean}$ and the heat transfer rate \qpl are constant for majority of the time, as demonstrated by the linearity of the melt fraction curve. Due to the Boussinesq approximation, the simulated flow conserves volume, and a downward movement of volume must be matched by an upward movement. Thus, even though we cannot explain the scaling entirely, we note that the conditions in the PCM match those given in the analysis of \citet[Eq 2.121]{Bejan2013}, which predicts a $Pr^{1/3}$ dependency. 

This implies that the temperature difference $T_{in}-T_{mean}$ is the most important parameter for obtaining fast heat transfer, and should be maximized. The HTF inlet temperature is constrained by the application being studied. Thus, the temperature difference may be maximized by picking a PCM with a lower mean melting temperature. However, increasing this difference corresponds to a loss in quality of heat stored. The temperature $T_{mean}$ is also expected to be an important parameter for heat transfer during discharging of the device, as it shall affect the discharging heat transfer rate. Thus, it is desirable to find an optimized value of $T_{mean}$ that maximizes the quality stored energy and the charging and discharging rates. 

\subsection{Effects of $Re_f$ and $Pr_f$: Discussion and implications}

Figure \ref{fig:Ref} shows \etatau for different values of $Re_f$. Increasing $Re_f$ 
reduces the boundary layer thickness in the HTF which increases $q_f$. We conclude 
that beyond $Re_f=800$, there is no significant improvement in $q_f$. Since the range 
of $Re_f$ presented here are in the laminar region, another possibility is that we 
would get further enhancement in $q_f$ with turbulent flow in the HTF pipe, which 
reduces the boundary layer thickness further. However, increasing the HTF thermal 
conductivity $k_f$ thus reducing the HTF Prandtl number $Pr_f$ also reduces the 
boundary layer thickness. Figure \ref{fig:Prp} shows that $q_f$ does not change 
significantly by increasing the thermal conductivity. This indicates that for the 
range of $Gr_p$ presented here, at $Re_f=800$, there is sufficient flow of energy into 
the HTF domain. Thus, it is desirable to maximize the energy available in the HTF region by selecting a HTF with sufficiently high conductivity and pumping it with sufficient velocity to remove the weak dependency of the heat transfer rate on $Re_f$ and $Pr_f$. As mentioned in \S\ref{sec:parametrizing_the_problem}, the fluid velocity is a parameter that can be easily controlled.

\section{Conclusions}
\label{sec:conclusions}
The parametric performance modeling of LHTES devices is essential for their
effective use. In this paper, we present a framework to analyse LHTES devices
and apply it to a typical shell and tube heat exchanger geometry. Out of all
the parameters listed in table \ref{tab:paralist}, we pick the fundamental
operating parameters $u_f$,$T_{in}$ and two fundamental material parameters $k_p$,$k_f$ and study their effect
on the heat transfer rate \qpl and the melt fraction $\eta$ by conducting a
matrix of 64 simulations. We observe that the melt fraction scales with the
PCM Grashof number $Gr_p^1$ and the PCM Prandtl number $Pr_p^{(1/3)}$ provided
that there is sufficient energy provided by the HTF. No significant scaling is
observed for the HTF Reynolds number $Re_f$ and HTF Prandtl number $Pr_f$ in
the range studied and we conclude that these parameters do not matter provided
that the heat transfer rate is not limited by the HTF.

The form of $\eta$ versus time as the PCM melts has a linear region and a
nonlinear region with the separation between them defined by a critical melt
fraction $\etac\approx 90\%$. The linear region is characterized by fast and
constant heat transfer rate which is a desired characteristic in LHTES
devices. The nonlinear region is characterized by an asymptotic approach to
fully melted and a corresponding asymptotic decrease in the heat transfer
rate. Contour plots of the liquid fraction at \etac for cases with vastly
different parameters are observed to be similar in shape, which suggests a
universality for the critical melt fraction \etac. Based on this, we make the
following conclusions about the design process for LHTES devices that shall
enable the maximization of heat transfer rates.

\begin{enumerate}
\item The HTF velocity and thermal conductivity have weak effects on
          the heat transfer rate, even at moderate values, provided that the
          HTF does not limit the overall availability of energy.
          As noted in \S\ref{sec:Results}, the
          velocity and the choice of HTF fluid are somewhat easier to
          customize than the PCM parameters, and the HTF velocity is limited
          only by considerations of optimizing pumping power and reducing pipe
          wear. This is termed as the `sufficient' condition, and is indicated
          by the HTF tube walls approaching constant temperature. The values
          $Re_f=800$ and $Pr_f=8$ are found to be sufficient for the geometry
          studied here.
	\item The variation of melt fraction $\eta$ (which is a measure of
          stored latent energy) with time consists of linear and asymptotic
          regions. The linear region is characterized by a constant and higher
          heat transfer rate, which makes it the relevant region for operating
          the heat exchanger as an energy storage device. The critical melt
          fraction $\etac$ denotes the transition between these regions, and
          the device should only be operated upto that value. For the current
          geometry, the value of \etac is 0.9.
	\item Given `sufficient' conditions in the HTF, the energy stored is
          given by the correlation $a Pr_p^{1/3} Gr_p Fo_p$ where $a$ is a
          constant governed by the particular choice of geometry. This
          correlation, applicable in the linear region, can also stated as a
          dimensionless timescale given in \eqref{eq:timeScaling}.
	\item The effect of
          the PCM Grashof number $Gr_p$ is much stronger than the PCM Prandtl
          number $Pr_p$. In terms of selecting PCM materials and operating
          parameters, this indicates that varying the melting point and/or HTF
          inlet temperature has a stronger effect on the heat transfer rate
          than enhancing the thermal conductivity of the PCM. However, if the
          charging and discharging HTF temperatures are fixed (this is
          expected, since they are governed by the application), increasing
          the charging $Gr_p$ reduces the discharging $Gr_p$. The melting
          point of the PCM should be optimized in order to satisfy both
          charging and discharging conditions. Hence, finding materials for
          which the melting point can be varied, with means such as additives
          or chemical composition, is indicated to be an important area for
          further research. This conclusion agrees with that of the the review
          presented by \citet{farid2004}
	\item The HTF Prandtl number is a parameter that can be used to eliminate limiting HTF effects on heat transfer. Thus, enhancement of HTF conductivity through additives is indicated as a future research subject.     
\end{enumerate}

\paragraph{Acknowledgments}

This work was sponsored by the US Department of Energy grants DE-EE00007708 and DE-EE00008277.

\clearpage
\bibliographystyle{plainnatnourl} 
\bibliography{bibliography}

\end{document}